\documentclass[preprint,aps,amsmath]{revtex4}
\usepackage{graphicx}
\newcommand{\bnabla}{\boldsymbol{\nabla}}

\begin{document}

\preprint{}

\title{Two-dimensional turbulence in magnetised plasmas}
\author{Alexander Kendl}
\affiliation{Institut f\"ur Ionenphysik und Angewandte Physik, Association
  Euratom-\"OAW,  Universit\"at Innsbruck, A-6020 Innsbruck, Austria
\vspace{1cm}}

\begin{abstract}
In an inhomogeneous magnetised plasma the transport of energy and particles
perpendicular to the magnetic field is in general mainly caused by
quasi two-dimensional turbulent fluid mixing.  
The physics of turbulence and structure formation is of ubiquitous
importance to every magnetically confined laboratory plasma for experimental or
industrial application. 
Specifically, high temperature plasmas for fusion energy research are
also dominated by the properties of this turbulent transport.  
Self-organisation of turbulent vortices to mesoscopic structures
like zonal flows is related to the formation of transport barriers that can
significantly enhance the confinement of a fusion plasma. 
This subject of great importance in research is rarely touched on in
introductory plasma physics or continuum dynamics courses. 
Here a brief tutorial on 2D fluid and plasma turbulence is presented as an
introduction to the field, appropriate for inclusion in undergraduate
and graduate courses. 

\vspace{6cm}

{\small \noindent \sl This is an author-created, un-copyedited version of an
  article published in European Journal of Physics 29, 911-926 (2008). IOP
  Publishing Ltd is not 
  responsible for any errors or omissions in this version of the manuscript or
  any version derived from it. The definitive publisher authenticated version
  is available online at doi: 10.1088/0143-0807/29/5/005.} 

\end{abstract}

\maketitle

\section{Introduction}

Turbulence is a state of spatio-temporal chaotic flow generically attainable
for fluids with access to a sufficient source of free energy. 
A result of turbulence is enhanced mixing of the fluid which is directed
towards a reduction of the free energy. 
Mixing typically occurs by formation of vortex structures on a large
range of spatial and temporal scales, that span between system,  energy
injection and dissipation scales \cite{frisch,davidson,sreenivasan}. 

Fluids comprise the states of matter of liquids, gases and plasmas
\cite{landau}.  
A common free energy source that can drive turbulence in neutral (or more
precisely: non-conducting) fluids is a strong enough gradient (or ``shear'')
in flow velocity, which can lead to vortex formation by Kelvin-Helmholtz
instability. Examples for turbulence occuring from this type of instability
are forced pipe flows, where a velocity shear layer is developing at the wall
boundary, or a fast jet streaming into a stationary fluid.
Another source of free energy is a thermal gradient in connection with an
aligned restoring force (as in liquids heated from below in a gravity field)
that leads to Rayleigh-Benard convection \cite{gallery}. 

Several routes for the transition from laminar flow to turbulence in fluids
have been proposed. 
For example, in some specific cases the Ruelle-Takens scenario occurs, where
by linear instability through a series of a few period doubling bifurcations
a final nonlinear transition to flow chaos is observed when a control
parameter (like the gradient of velocity or temperature) is increased
\cite{ott}. For other scenarios, like in pipe flow, a sudden direct
transition by subcritical instability to fully developed turbulence or an
intermittent transition are possible \cite{grossmann00,drazin}. 

The complexity of the flow dynamics is considerably enhanced in a plasma
compared to a non-conducting fluid. A plasma is a macroscopically neutral gas
composed of many electrically charged particles that is essentially determined
by collective degrees of freedom  \cite{chen,spatschek}. 
Space and laboratory plasmas are usually composed
of positively charged ions and negatively charged electrons that are
dynamically coupled by electromagnetic forces.
Thermodynamic properties are governed by collisional equilibration and
conservation laws like in non-conducting fluids. 
The additional long-range collective interaction by spatially and temporally
varying electric and magnetic fields allows for rich dynamical behaviour of
plasmas with the possibility for complex flows and structure formation in the
presence of various additional free energy sources \cite{horton}. 

The basic physics of plasmas in space, laboratory and fusion experiments is
introduced in detail in a variety of textbooks (e.g. in
Refs.~\cite{cap,goldston,sturrock}).  

Although the dynamical equations for fluid and plasma flows can be
conceptually simple, they are highly nonlinear and involve an infinite number
of degrees of freedom \cite{biskamp}. 
Analytical solutions are therefore in general
impossible. The description of fluid and plasma dynamics mainly relies on
statistical and numerical methods. 

\begin{figure}
\includegraphics[width=7.0cm]{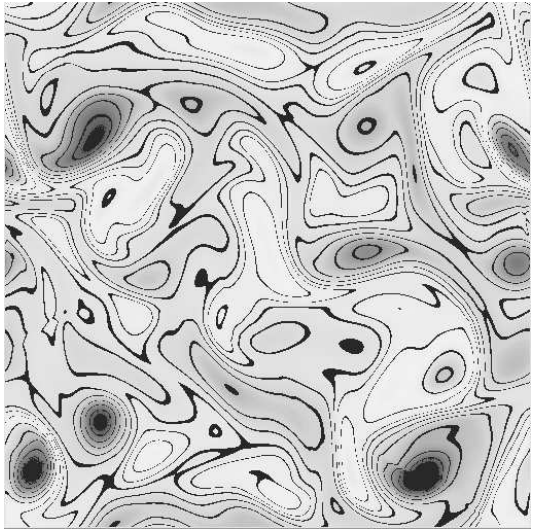}
\caption{\label{f:decay} \sl
Computation of decaying two-dimensional fluid turbulence, showing contours\\
of vorticity ${\boldsymbol  \omega} = \bnabla \times {\bf u}$.
}
\end{figure}

\section{Continuum dynamical theory of fluids and plasmas}

Computational models for fluid and plasma dynamics may be broadly classified
into three major categories:

\begin{itemize}
\item 
(1) Microscopic models: many body dynamical description by ordinary
  differential equations and laws of motion;
\item 
(2) Mesoscopic models: statistical physics description (usually by
  integro-differential equations) based on probability
  theory and stochastic processes;
\item 
(3) Macroscopic models: continuum description by partial differential
  equations based on conservation laws for the
  distribution function or its fluid moments.
\end{itemize}

Examples of microscopic models are Molecular Dynamics (MD) methods for neutral
fluids that model motion of many particles connected by short range
interactions \cite{rapaport}, or Particle-In-Cell (PIC) methods for plasmas
including electromagnetic forces \cite{birdsall,particlesim05}. 
Such methods become important when relevant
microscopic effects are not covered by the averaging procedures used to obtain
meso- or macroscopic models, but they usually are intrinsical computationally
expensive. 

Complications result from multi-particle or multi-scale interactions. 
Mesoscopic modelling treats such effects on the dynamical evolution of
particles (or modes) by statistical assumptions on the interactions
\cite{stochastic}. 

These may be implemented either on the macroscale as spectral fluid
closure schemes like, for example, in the Direct Interaction Approximation
(DIA), or on the microscale as advanced collision operators like in
Fokker-Planck models. An example of a mesoscopic computational model for fluid
flow is the Lattice Boltzmann Method (LBM) that combines free streaming
particle motion by a minimalistic discretisation in velocity space with
suitable models for collision operators in such a way that fluid motion is
recovered on macroscopic scales. 

Macroscopic models are based on the continuum description of the kinetic
distribution function of particles in a fluid, or of its hydrodynamic moments. 
The continuum modelling of fluids and plasmas is introduced in more detail
below. 

Computational methods for turbulence simulations have been developed within
the framework of all particle, mesoscopic or continuum models. Each of the
models has both advantages and disadvantages in their practical numerical
application. The continuum approach can be used in situations where discrete
particle effects on turbulent convection processes are negligible. 
This is to some approximation also the case for many situations and regimes
of interest in fusion plasma experiments that are dominated by turbulent
convective transport, in particular at the (more collisional) plasma edge. 

Within the field of Computational Fluid Dynamics the longest experience
and broadest applications have been obtained with continuum methods
\cite{wesseling}. 
Many numerical continuum methods that were originally developed for neutral
fluid simulation have been straightforwardly applied to plasma physics
problems \cite{tajima}. 

In continuum kinetics, the time evolution of the single-particle probability
  distribution function $f({\bf x}, {\bf v}, t)$ for particles of each species
  (e.g. electrons and ions in a plasma) in the presence of a mean
  force field ${\bf  F}({\bf x}, t)$ and within the binary collision
  approximation (modelled by an operator $C$) is described by the Boltzmann
  equation \cite{boltzmann}  
\begin{equation}
\left( \partial_t  + {\bf v} \cdot \partial_{\bf x}  + {\bf F} \cdot
\partial_{\bf v} \right) f = C f.
\end{equation}
In a plasma the force field has to be self-consistently determined by
solution of the Maxwell equations. Usually, kinetic theory and computation for
gas and plasma dynamics make use of further simplifying approximations that
considerably reduce the complexity: in the Vlasov equation binary collisions
are neglected ($C=0$), and in the drift-kinetic or gyro-kinetic plasma
equations further reducing assumptions are taken about the time and space
scales under consideration. 

The continuum description is further simplified when the fluid can be assumed
to be in local thermodynamic equilibrium. Then a hierarchical set of
hydrodynamic conservation equations is obtained by construction of moments
over velocity space \cite{chapman,sone}. 
In lowest orders of the infinite hierarchy, the
conservation equations for mass density $n({\bf x}, t)$, momentum $n {\bf u}
({\bf x}, t)$ and energy density ${\cal E} ({\bf x}, t)$ are obtained. Any
truncation of the hierarchy of moments requires the use of a closure scheme
that relates quantities depending on higher order moments by a constitutive
relation to the lower order field quantities. 

An example of a continuum model for neutral fluid flow are the Navier-Stokes
 equations. In their most widely used form (in particular for technical and
 engineering applications) the assumptions of incompressible
divergence free flow (i.e., $n$ is constant on particle paths) and of an
 isothermal equation of state are taken \cite{foias}. 

Then the description of fluid flow can be reduced to the solution of the
(momentum) Navier-Stokes equation 
\begin{equation}
\left( \partial_t + {\bf u} \cdot \bnabla \right) {\bf u} = 
- \bnabla \! P + \nu \Delta {\bf u}
\label{e:nse}
\end{equation}
under the constraints given by 
\begin{equation}
\nabla \cdot {\bf u} \equiv 0
\end{equation}
and by boundary
  conditions. Most numerical schemes for the 
  Navier-Stokes equation require solution of a Poisson type equation for the
 normalised scalar pressure $P = p/\rho_0$ in order to guarantee divergence
  free flow.  

The character of solutions for the Navier-Stokes equation intrinsically
depends on the ratio between the dissipation time scale (determined by the
kinematic viscosity ${\nu}$) and the mean flow time scale (determined by the
system size $L$ and mean velocity $U$), specified by the Reynolds number
\begin{equation}
R_e = { L U / \nu }.
\end{equation}
For small values of $R_e$ the viscosity will dominate the time evolution of
${\bf u}({\bf x})$ in the Navier-Stokes equation, and the flow is
laminar. For higher $R_e$ the advective nonlinearity is dominant and the
flow can become turbulent. 
The Rayleigh number has a similar role for the onset of
thermal convective turbulence \cite{ott}.

\section{Drift-reduced two-fluid equations for plasma dynamics}

Flow instabilities as a cause for turbulence, like those driven by flow shear
or thermal convection, do in principle also exist in plasmas similar to
neutral fluids \cite{biskamp-turb}, but are in general found to be less
dominant in strongly magnetised plasmas. 
The most important mechanism which results in turbulent transport and enhanced
mixing relevant to confinement in magnetised plasmas
\cite{hasmim78,horton81,wakatani84} is an 
unstable growth of coupled wave-like perturbations in plasma pressure
$\tilde p$ and electric fields $\tilde {\bf E}$. 
The electric field forces a flow with the ExB (``E-cross-B'') drift velocity 
\begin{equation}
{\bf v}_{ExB} = {1 \over B^2} \tilde {\bf E} \times {\bf B} 
\end{equation}
of the plasma
perpendicular to the magnetic field {\bf B}. A phase shift, caused by any
inhibition of a fast parallel Boltzmann response of the electrons,
between pressure and electric field perturbation in the presence of a pressure
gradient can lead to an effective transport of plasma across the magnetic
field and to an unstable growth of the perturbation amplitude.  
Nonlinear self-advection of the ExB flow and coupling between perturbation
modes (``drift waves'') can finally lead to a fully developed turbulent state
with strongly enhanced mixing.  

A generic source of free energy for magnetised laboratory plasma turbulence
resides in the pressure gradient: in the core of a magnetic confinement region
both plasma density and temperature are usually much larger than near the
bounding material wall, resulting in a pressure gradient directed inwards to
the plasma center. 
Instabilities which tap this free energy tend to lead to enhanced mixing and
transport of energy and particles down the gradient \cite{lnp-stroth}. 
For magnetically confined fusion plasmas, this turbulent convection by
ExB drift waves often dominates collisional diffusive transport mechanisms
by orders of magnitude, and essentially determines energy and particle
confinement properties \cite{itoh99,hinton76}.  
The drift wave turbulence is regulated by formation of mesoscopic streamers
and zonal structures out of the turbulent flows \cite{zf-review05}. 

Continuum models for drift wave turbulence have to capture the
different dynamics of electrons and ions parallel and perpendicular to the
magnetic field and the coupling between both species by electric and
magnetic interactions \cite{itoh01,itoh03}. 
Therefore, a single-fluid magneto-hydrodynamic (MHD)
model can not appropriately describe drift wave dynamics: one has to refer to
a set of two-fluid equations, treating electrons and ions as separate species,
although the plasma on macroscopic scales remains quasi-neutral with nearly
identical ion and electron density, $n_i \approx n_e \equiv n$.

The two-fluid equations require quantities like collisional momentum exchange
rate, pressure tensor and heat flux to be expressed by hydrodynamic moments
based on solution of a kinetic (Fokker-Planck) model. The most widely used set
of such fluid equations has been derived by Braginskii \cite{braginskii} and
is e.g. presented in brief in Ref.~\cite{wesson}.

The most general continuum descriptions for the plasma
species, based either on the kinetic Boltzmann equation or on the hydrodynamic
moment approach like in the Braginskii equations, are covering all time and
space scales, including detailed 
gyro-motion of particles around the magnetic field lines, and the fast plasma
frequency oscillations. From experimental observation it is on the other hand
evident \cite{hugill83,liewer85,wootton90,wagner93,stroth98}, that the
dominant contributions to turbulence and transport in 
magnetised plasmas originate from time and space scales that are associated
with frequencies in the order the drift frequency $\omega \sim (\rho_s /
L_{\perp}) \Omega_i$, that are much lower than the ion gyro-frequency
$\Omega_i = q_i B/M_i$ by the ratio  between drift scale $\rho_s = \sqrt{T_e
  M_i}/(eB)$ to gradient length $L_{\perp}$:
\begin{equation}
\omega \sim \partial_t \ll \Omega_i < \Omega_e.
\end{equation}
Under these assumptions one can apply a ``drift ordering'' based on
the smallness of the order parameter $\delta = \omega / \Omega_i \ll 1$. 
This can be introduced either on the kinetic level, resulting in the
drift-kinetic model, or on the level of two-fluid 
moment equations for the plasma, resulting in the ``drift-reduced two-fluid
equations'', or simply called ``drift wave equations'' 
\cite{hinton71,tang78,wakatani84}:
neglect of terms scaling with $\delta$ in
higher powers than 2 considerably simplifies both the numerical
and analytical treatment of the dynamics, while retaining all effects of the
perpendicular drift motion of guiding centers and nonlinear couplings
that are necessary to describe drift wave turbulence. 
 
For finite ion temperature, the ion gyro-radius $\rho_i = \sqrt{T_i M_i}/(eB)$
can be of the same magnitude as typical fluctuation scales, with wave numbers
found around $k_{\perp} \rho_s \sim 0.3$ in the order of the drift scale
\begin{equation}
\rho_s = {\sqrt{T_e M_i} \over eB}.  
\end{equation}
Although the gyro-motion is still fast compared to turbulent time scales, the
ion orbit then is of similar size as spatial variations of the fluctuating
electrostatic potential. 
Finite gyro-radius (or ``finite Larmor radius'', FLR) effects are captured by
appropriate averaging procedures over the gyrating particle trajectory and
modification of the polarisation equation, resulting in ``gyrokinetic'' or
``gyrofluid models'' for the plasma.

\section{Turbulent vortices and mean flows}

The prevalent picture of drift wave turbulence is that of small-scale,
low-frequency ExB vortices in the size of several gyro-radii, that determine
mixing and transport of the plasma perpendicular to the magnetic field across
these scales. 

Beyond that, turbulence in magnetised plasmas exhibits large-scale structure
formation that is linked to this small-scale eddy motion: 
The genesis of mean zonal flow structures out of an initially nearly
homogeneous isotropic vortex field and the resulting shear-flow suppression of
the driving turbulence is a particular example of a self-organising regulation
process in a dynamical system
\cite{hasmim78,haswak87,biglari90,lin98,terry00rmp,terry00pop,hahm00,hahm02}.  
The scale of these macroscopic turbulent zonal flows is that of the system
size, setting up a radially localised differential ExB rotation of the whole
plasma on an entire toroidal flux surface.

\begin{figure}
\includegraphics[width=15.0cm]{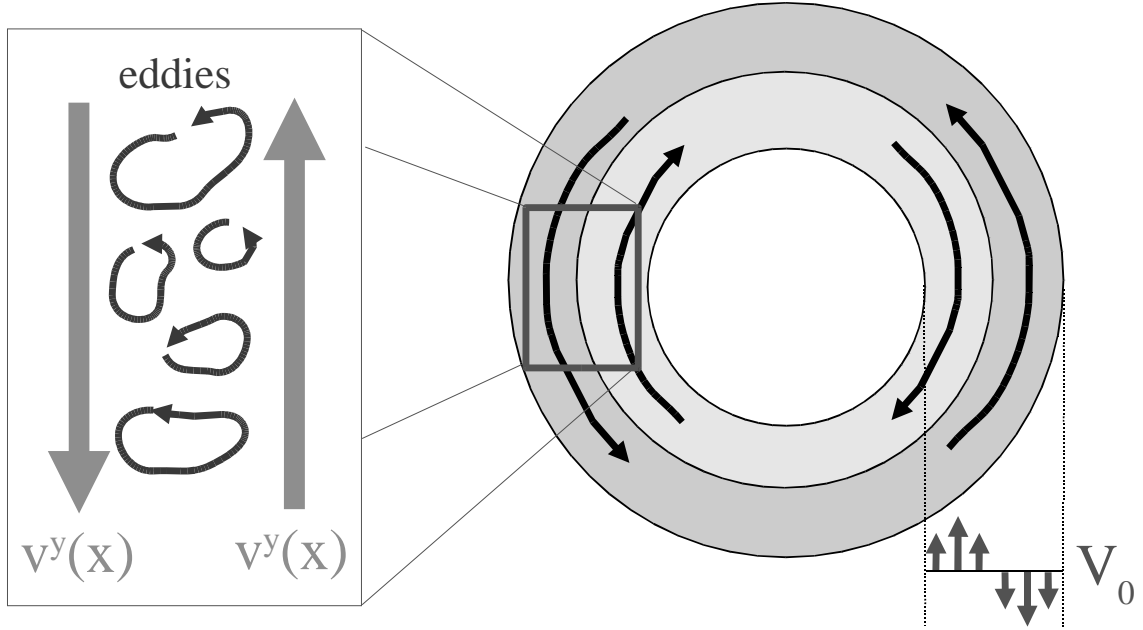}
\caption{\label{f:eddy-flow} \sl Generation of mean sheared flows from drift wave
  turbulence within the poloidal cross-section of a magnetised plasma torus.}
\end{figure}

Moreover, the process of self-organisation to zonal flow structures is thought
to be a main ingredient in the still unresolved nature of the L-H transition
in magnetically confined toroidal plasmas for fusion research \cite{connor00}. 
The L-H transition is experimentally found to be a sudden improvement
in the global energy content of the fusion plasma from a low to high (L-H)
confinement state when the central heat source power is increased above a
certain threshold \cite{wagner82,gohil94,hugill00,suttrop97}. 
The prospect of operation in a high confinement
H-mode is one of the main requirements for successful performance of fusion
experiments like ITER.

The mechanism for spin-up of zonal flows in drift wave turbulence
is a result of the quasi two-dimensional nature of the nonlinear ExB dynamics,
in connection with the double periodicity and parallel coupling in a toroidal
magnetised plasma. 

Basic concepts and terminology for the interaction between vortex turbulence
and mean flows have been first developed in the context of neutral fluids.
It is therefore instructive to briefly review these relations in the framework
of the Navier-Stokes Eq.~(\ref{e:nse}) before applying them to plasma dynamics.

Small (space and/or time) scale vortices and mean flows may be separated
formally by an ansatz known as Reynolds decomposition,
\begin{equation}
{\bf u} = \bar{ \bf U} + \tilde {\bf u},
\end{equation}
splitting the flow velocity into a mean part $\bar {\bf U} = \langle {\bf u}
\rangle$, averaged over the separation scale, and small-scale fluctuations
$\tilde {\bf u}$ with $\langle \tilde {\bf u}\rangle =0$. While the averaging
procedure, $\langle ...\rangle$, is mathematically most unambiguous for the
ensemble average, the physical interpretation in fluid dynamics makes a time
or space decomposition more appropriate.  
Applying this averaging on the Navier-Stokes Eq.~(\ref{e:nse}), one obtains the
Reynolds equation (or: Reynolds averaged Navier-Stokes equation, RANS):
\begin{equation}
\left( \partial_t + \bar{\bf U} \cdot \bnabla \right) \bar{\bf U} = 
- \bnabla \! \bar{P} + \bnabla {\bf R} + \nu \Delta \bar{\bf U}
\label{e:rans}
\end{equation}
This mean flow equation has the same structure as the original Navier-Stokes
equation with one additional term including the Reynolds stress tensor $R_{ij}
= \langle \tilde u_i \tilde u_j \rangle$. Momentum transport between turbulence
and mean flows can thus be caused by a mean pressure gradient, viscous forces,
and Reynolds stress. A practical application of the RANS is in Large Eddy
Simulation (LES) of fluid turbulence, which efficiently reduces the time and
space scales necessary for computation by {\sl modelling} the Reynolds stress
tensor for the smaller scales as a local function of the large scale flow. 
LES is however not applicable for drift wave turbulence computations, as
here in any case all scales down to the effective gyro-radius (or drift scale
$\rho_s$) have to be resolved in Direct Numerical Simulation (DNS). 

\section{Two-dimensional fluid turbulence}

Turbulent flows are generically three-dimensional. In some particular
situations the dependence of the convective flow dynamics on one of the
Cartesian directions can be negligible compared to the others and the
turbulence becomes quasi two-dimensional \cite{kraichnan80,tabeling02}.  
Examples for such 2D fluid systems are thin films
(e.g. soap films), rapidly rotating stratified fluids, or geophysical flows of
ocean currents and the (thin) planetary atmosphere. In particular, also the
perpendicular ExB dynamics in magnetised plasmas behaves similar to a 2D
fluid \cite{horton94}.  
The two-dimensional approximation of fluid dynamics not only simplifies the
treatment, but moreover introduces distinctly different behaviour. 

The major difference can be discerned by introducing the vorticity 
\begin{equation}
{\boldsymbol  \omega} = \bnabla \times {\bf u} 
\end{equation}
and taking the curl of the Navier-Stokes Eq.~(\ref{e:nse})
to get the vorticity equation
\begin{equation}
\left( \partial_t + {\bf u} \cdot {\bnabla} \right) {\boldsymbol \omega} = 
\left( {\boldsymbol \omega} \cdot {\bnabla} \right) {\bf u} + \nu \Delta {\boldsymbol
  \omega}. 
\label{e:voreq}
\end{equation}
In a two-dimensional fluid with ${\bf v} = v_x {\bf e}_x + v_y {\bf e}_y$ and
$v_z = 0$ the vorticity reduces to ${\boldsymbol \omega} = w {\bf e}_z$ with
$w= \partial_x v_y - \partial_y v_x$. 
The vortex stretching and twisting term $\left( {\boldsymbol \omega} \cdot
{\bnabla} \right) {\bf u}$ is zero in 2D, thus eliminating a characteristic
feature of 3D turbulence. 
For unforced inviscid 2D fluids then due to $\left( \partial_t + {\bf u}
\cdot {\bnabla}\right) {\boldsymbol \omega} = 0$ the vorticity $w$ is constant
in  flows along the fluid element. This implies conservation of total enstrophy
$W = \int (1/2) |{\boldsymbol \omega}|^2 d{\bf x}$ in addition to the
conservation of kinetic flow energy $E = \int (1/2) |{\bf u}|^2 d{\bf x}$. 

The 2D vorticity equation can be further rewritten in terms of a scalar stream
function $\phi$ that is defined by $(v_x, v_y) = (\partial_y, - \partial_x)
\phi$ so that $w = \nabla^2 \phi$, to obtain
\begin{equation}
\partial_t w  + \left[ \phi, w \right] = \nu \Delta w.
\label{e:vor2d}
\end{equation}
Here the Poisson bracket $[a,b] = \partial_y a \; \partial_x b - \partial_x a
\; \partial_y b$ is introduced. For force driven flows a term given
  by the curl of the force adds to the right hand side of Eq.~(\ref{e:vor2d}).
Although the pressure is effectively eliminated from that equation, it is
still necessary to similarly solve a (nonlocal) Poisson equation for the
stream function. For ExB flows in magnetised plasmas the stream function
$\phi$ is actually represented by the fluctuating electrostatic potential.

The energetics of homogeneous 3D turbulence is usually understood in terms
of a direct cascade of energy from the injection scale down to small
(molecular) dissipation scales \cite{frisch}: large vortices break up into
smaller ones due to mutual stretching and shearing. 
In terms of the Reynolds Eq.~(\ref{e:rans}) this means that the Reynolds stress
transfer is usually negative, taking energy out of mean flows into
small scale vortices.
The interaction between scales takes place basically by
three-mode coupling maintained by the convective quadratic nonlinearity.
This leads to the generic Kolmogorov $k^{-5/3}$ power spectrum of Fourier
components $E(k) = \int dx \; (1/2) |{\bf u}(x)|^2 \exp(-ikx)$
in the cascade range of 3D turbulence when energy injection and dissipation
scales are well separated by a high Reynolds number \cite{k41}. 

In two dimensions the behaviour is somewhat different: Kraichnan, Leith and
Batchelor \cite {kraichnan67,leith68,batchelor69} have conjectured that the
energy has an inverse cascade property to scales larger than the injection. 
Smaller vortex structures self-organise to merge into bigger ones as a result
of the absence of vortex stretching. For unforced turbulence the Reynolds
stress transfer is on the average positive and into the mean flows.  

The classical theory of 2D fluid turbulence by Kraichnan et al. predicts a
$k^{-3}$ energy spectrum and a $k^{-1}$ enstrophy spectrum in the inertial
range. 
Numerical simulations of 2D Navier-Stokes turbulence however rather find, for
example, 
a $k^{-5/3}$ inverse cascade for energy on large scales and a $k^{-3}$ direct
cascade for enstrophy on small scales \cite{tran03}, modifying the classical
predictions due to the existence of intermittency and coherent structures,
although the extent of the modification is still under discussion. The
(limiting or periodic) domain boundary in 2D simulations has also been found
to have stronger influence than for the 3D case.  

Periodicity in one dimension of the 2D system can lead to the spin up of
sustained zonal structures of the mean flow out of the turbulence by inverse
cascade.  
Prominent examples of zonal flows in planetary atmospheric dynamics are the
well visible structures spanning around the planet Jupiter approximately along
constant latitude, and jet streams in the earth's atmosphere. Zonal flows are
also observed in fluids rotating in a circular basin.

Drift wave turbulence in magnetised plasmas also has basically a 2D character
and exhibits zonal structure formation in the poloidally and toroidally
periodic domain on magnetic flux surfaces of a torus. These zonal plasma flows
have finite radial extension and constitute a differential, sheared rotation
of the whole plasma on flux surfaces. 

\begin{figure}
\includegraphics[width=16.5cm]{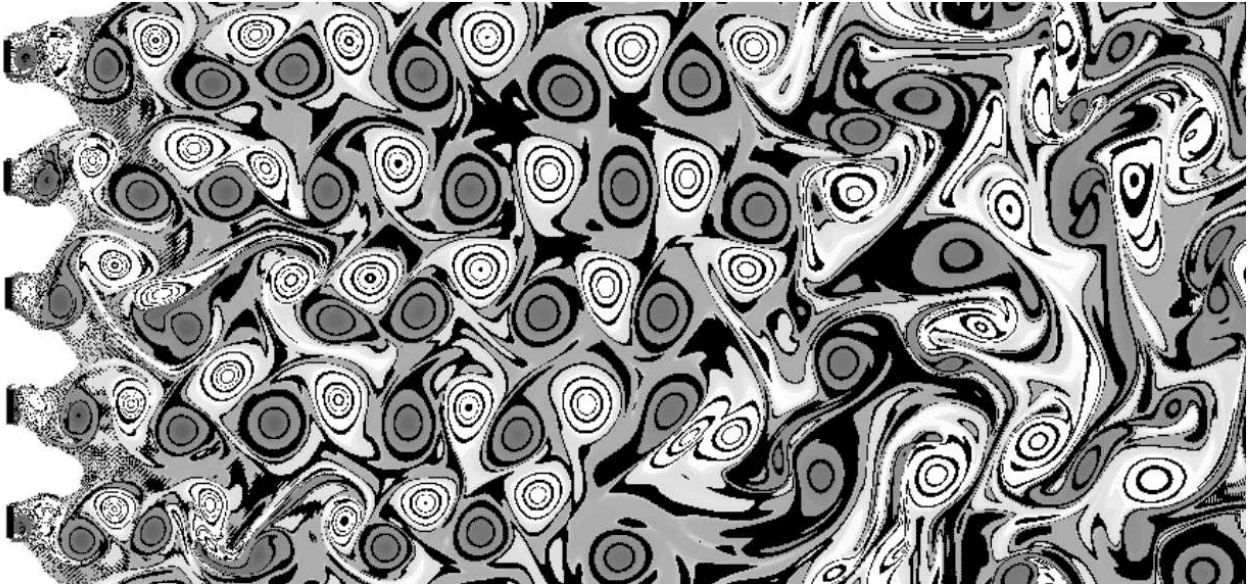}
\caption{\label{f:lbm} \sl
Example for a fluid simulation of 2D grid turbulence with a Lattice-Boltzmann
code~\cite{C45}: a high Reynolds number flow with $R_e=5000$ is entering from
the left of the domain and passes around a grid of obstacles. The shading
depicts vorticity. In the near field directly behind the grid the particular
vortex streets can be distinguished. In the middle of the domain neighbouring
eddies are strongly coupled to a quasi homogeneous (statistically in the
perpendicular direction) vortex field. On the far right side eddies decay
into larger structures in the characteristic way of 2D turbulence. The
simulation agrees well with flowing soap film experiments~\cite{soapfilm}.
}
\end{figure}

\section{Turbulence in magnetised plasmas}

Drift wave turbulence is nonlinear, non-periodic motion involving
disturbances on a background thermal gradient of a magnetised plasma
and eddies of fluid like motion in which the advecting velocity of all
charged species is the ExB velocity. The
disturbances in the electric field ${\bf E}$ implied by the presence
of these eddies are caused by the tendency of the electron dynamics to
establish a force balance along the magnetic field ${\bf B}$. 

Pressure disturbances have their parallel gradients balanced by a parallel
electric field, whose static part is given by the parallel gradient of
the electrostatic potential. This potential in turn is the stream
function for the ExB velocity in drift planes,
which are locally perpendicular to the magnetic field. The turbulence
is driven by the background gradient, and the electron pressure and
electrostatic potential are coupled together through parallel
currents. Departures from the static force balance are mediated
primarily through electromagnetic induction and resistive friction,
but also the electrons inertia, which is not negligible.

The dynamical character of cross-field ExB drift wave turbulence in the edge
region of a tokamak plasma is governed by this electromagnetic and dissipative
effects in the parallel response.  

The most basic drift-Alfv\'en (DALF) model to capture the drift wave
dynamics
includes nonlinear evolution equations of three fluctuating fields: the
electrostatic potential $\tilde \phi$, electromagnetic potential $\tilde
A_{||}$, and density $\tilde n$. The tokamak edge usually features a more or
less pronounced density pedestal, and the dominant contribution to the free
energy drive to the turbulence by the inhomogeneous pressure background is
thus due to the density gradient. 

On the other hand, a steep enough ion
temperature gradient (ITG) does not only change the turbulent
transport quantitatively, but adds new interchange physics into the
dynamics. In addition, more field quantities have to be treated:
parallel and perpendicular temperatures $\tilde T_\parallel$ and $\tilde
T_\perp$ and the associated parallel heat fluxes, for a total of six
moment variables for each species. Finite Larmor radius effects
introduced by
warm ions require a gyrofluid description of the turbulence equations.  

Both the resistive DALF and the ITG models can be covered by using the
six-moment electromagnetic gyrofluid model GEM by Scott \cite{gem}, but for
basic studies it is also widely used in its more economical two-moment version
for scenarios where the DALF model is applicable \cite{scott03ppcf}.
The gyrofluid model is based upon a moment approximation of the underlying
gyrokinetic equation.  

The first complete six-moment gyrofluid formulation was given for slab
geometry by Dorland et al. \cite{dorland93}, and later extended by Beer et
al. to incorporate toroidal effects \cite{beer96a} using a ballooning-based
form of flux surface geometry \cite{beer95}. 

Electromagnetic induction and electron collisionality were then included
to form a more general gyrofluid for edge turbulence by Scott \cite{scott00},
with the geometry correspondingly replaced by the version from the edge
turbulence work, which does not make ballooning assumptions and in
particular represents slab and toroidal mode types equally well and does
not require radial periodicity \cite{scott01}.  Energy conservation
considerations were solidified first for the two-moment version
\cite{scott03ppcf}, and recently for the six-moment version in
Ref.~\cite{gem}.  
 
\begin{figure}
\includegraphics[width=16.5cm]{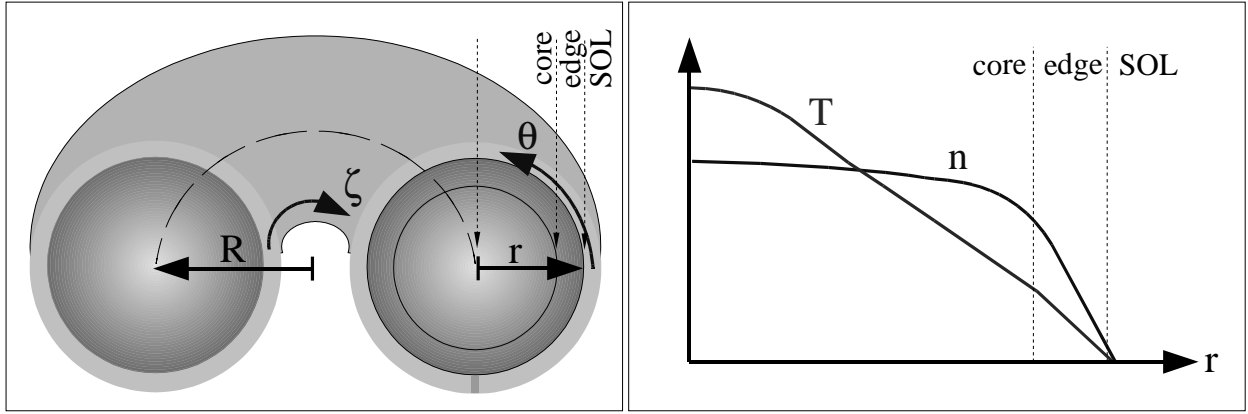}
\caption{\label{f:profile} \sl
Left: A cut of a torus shows the poloidal cross section with minor
radius $r$, major radius $R$, poloidal angle $\theta$ and toroidal angle
$\zeta$. Right: typical radial profiles of density $n(r)$ and temperature
$T(r)$ in the toroidal plasma of a tokamak. The density often is nearly
constant in the ``core'' region and shows a pronounced steep-gradient
pedestal in the plasma ``edge'' region. The outer scrape-off layer (``SOL'')
connects the plasma with materials walls.
}
\end{figure}

\section{Basic drift wave instability}

Destabilization of the ExB drift waves occurs when the parallel electron
dynamics deviates from a fast ``adiabatic'' response to potential
perturbations, resulting in a phase shift between the density and potential
fluctuations. 
In this section the basic linear instability mechanism is discussed in the
most basic electrostatic, cold ion limit for a straight magnetic field.

\begin{figure}
\includegraphics[width=16.5cm]{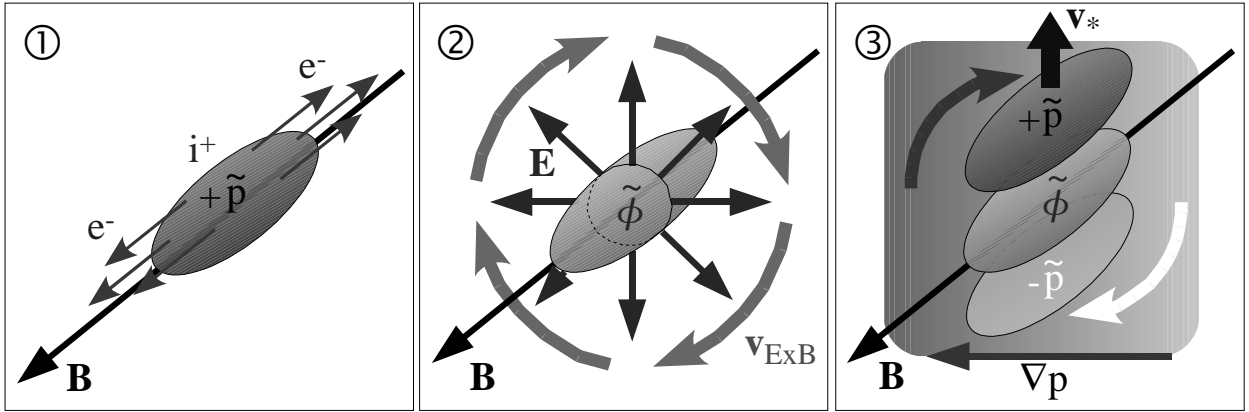}
\caption{\label{f:dw} \sl
Basic drift waves mechanism: (1) an initial pressure perturbation
$\tilde p$ leads to an ambipolar loss of electrons along the magnetic
field ${\bf B}$, whereas ions remain more immobile. (2)
The resulting electric field ${\bf \tilde E}=-\bnabla \tilde \phi$ convects
the whole plasma 
with ${\bf v}_{ExB}$ around the perturbation in the plane perpendicular to
${\bf B}$. (3) In the presence of a pressure gradient, $\tilde p$
propagates in electron diamagnetic
drift direction with ${\bf v}_{\ast} \sim \bnabla p \times {\bf B}$. 
This ``drift wave'' is stable if 
the electrons establish $\tilde \phi$ according to the Boltzmann relation
without delay (``adiabatic response''). 
A non-adiabatic response due to collisions, magnetic flutter or wave-kinetic
effects causes a phase shift between $\tilde p$ and $\tilde \phi$.
The ExB velocity is then effectively transporting plasma down
the gradient, enhances the principal perturbation and leads to an
unstable growth of the wave amplitude
}
\end{figure}

Figure~\ref{f:dw} schematically shows a localized perturbation
of plasma pressure $\tilde p$ (left) that results in a positive potential
perturbation $\tilde \phi>0$ (middle) due to ambipolar diffusion. 
For typical tokamak parameters it is found that the perturbation scale 
$\Delta \gg \lambda_D = \sqrt{\varepsilon_0 T_e / (ne^2)}$ is much larger than
the Debye length $\lambda_D$, so that quasi neutrality $n_i \approx n_e \equiv
n$ can be assumed. 

In accordance with the stationary parallel electron momentum balance equation 
\begin{equation}
-en_0{\tilde {\bf E}}_{||} - \bnabla_{||} \tilde p_e=0
\label{e:emom}
\end{equation}
with $\tilde p_e = \tilde n_e T_e$, the isothermal electrons try to locally
establish  along the field line
a Boltzmann relation  $n_e = n_0(r) \exp( e \tilde \phi / T_e)$.
Under quasi neutrality $n_i$=$n_e$=$n_0(r)$+$\tilde n_e$, where $n_0(r)$ is
the (in general radially varying) background density.

Without restrictions on the parallel electron dynamics (like e.g. due
to collisions, Alfv\'en waves or kinetic effects like Landau damping
and particles trapped in magnetic field inhomogeneities) this balance
is established instantaneously on the drift time scale and is usually
termed an ``adiabatic response''.

Already at homogeneous background density the perturbation convects the
plasma with the ExB drift velocity ${\bf v}_{\perp} = {\bf v}_{ExB} = (B^{-2})
\tilde {\bf E} \times {\bf B}$ equal for electrons and ions. 
When a perpendicular background pressure gradient $\bnabla p$ is present, the
perturbed structure propagates in the electron diamagnetic drift direction
$\sim \bnabla p \times {\bf B}$.  

In the continuity equation
\begin{equation}
\partial_t n + \bnabla \cdot (n {\bf v}) = 0 
\label{e:cont}
\end{equation}
for cold ions in a homogeneous magnetic field and by neglecting ion inertia the
only contribution to the velocity is the perpendicular ExB drift velocity
${\bf v}_E = - B^{-2} (\nabla \tilde \phi \times {\bf B})$. Using the
Boltzmann relation in Eq.~(\ref{e:cont}) one gets
\begin{equation}
\partial_t \; n_0 \exp \left( {e \tilde \phi \over T_e} \right) - \bnabla \cdot
\left[
{1 \over B^2} (\nabla \tilde \phi  \times {\bf B}) \: n_0 \exp \left({e \tilde
  \phi \over T_e} \right) \right] = 0, 
\end{equation}
and due to the straight ${\bf B}=B {\bf e}_{\parallel}$ it is obtained:
\begin{equation}
\partial_t \tilde \phi - \left( {T_e \over e B} \right) (\partial_r \ln n_0)
\; \partial_{\theta} \tilde \phi = 0.
\end{equation}

Assuming a perturbation periodical in the electron diamagnetic drift coordinate
$\theta$ with $\tilde \phi = \tilde \Phi \exp[-i \omega t + i k_{\theta}
  \theta]$, the electron drift wave frequency is found to be 
\begin{equation}
\omega_{\ast e} = {T_e \over eB} {1 \over L_n} k_{\theta}
=
{c_s \over L_n} [\rho_s k_{\theta}] 
= {\rho_s \over L_n} \Omega_i [\rho_s k_{\theta}].
\end{equation}
Here the density gradient length $L_n = (\partial_r \ln n_0)^{-1}$
and the drift scale $\rho_s= \sqrt{m_i  T_e}/(eB)$, representing an ion
radius at electron temperature, have been introduced.

The motion of the perturbed structure perpendicular to magnetic field and
pressure gradient in the electron diamagnetic drift direction $k_{\theta} \sim 
\bnabla p \times {\bf B}$ is in this approximation still stable and does so far
not cause any transport down the gradient.

The drift wave is destabilized only when a phase shift $\delta_{\bf
k}$ between potential and density perturbation is introduced by
``non-adiabatic'' electron dynamics 
\begin{equation}
\tilde n_e = n_0 (1-i\delta_{\bf k}) \: {e \tilde \phi \over T_e}
\label{e:nonadiab}
\end{equation}
The imaginary term $i\delta_{\bf k}$ in general is an anti-hermitian
operator and describes dissipation of the electrons, that causes the
density perturbations to proceed the potential perturbations in $\theta$ by
slowing down the parallel equilibration. This leads to an exponential
growth of the perturbation amplitude by $\exp (\gamma_k t)$ with linear growth
rate $\gamma_k \sim \delta_k \omega_k$. 

Parallel electron motion also couples drift waves to shear Alfv\'en
waves, which are parallel propagating perpendicular magnetic field
perturbations. With the vector potential $A_{||}$ as a further dynamic
variable, the parallel electric field $E_{||}$, parallel electron
motion, and nonlinearly the parallel gradient are modified.
The resulting nonlinear drift-Alfv\'en equations are discussed in the following
section.

The stability and characteristics of drift waves and resulting plasma
turbulence are further influenced by inhomogeneities in the magnetic field, in
particular by field line curvature and shear. 
The normal and geodesic components of field line curvature have different
roles for drift wave turbulence instabilities and saturation. 
The field gradient force associated with the normal curvature, if
aligned with the plasma pressure gradient, can either act to restore or amplify
pressure gradient driven instabilities by compression of the fluid drifts,
depending on the sign of alignment. 
The geodesic curvature describes the compression of the field strength in
perpendicular direction on a flux surface and is consequently related to the
compression of large-scale (zonal) ExB flows.

Transition from stable drift waves to turbulence has been studied
experimentally in linear and simple toroidal magnetic field configurations,
and by direct numerical simulation. 

Experimental investigations in a magnetized low-beta plasma with clindrical
geometry by Klinger {\sl et al.} have demonstrated that the spatiotemporal
dynamics of interacting destabilised travelling drift wave follows a
bifurcation sequence towards weakly developed turbulence according to the
Ruelle-Takens-Newhouse scenario \cite{klinger97}.  
The relationship between observations made in linear magnetic geometry,
purely toroidal geometry and magnetic confinement is discussed in
Ref.~\cite{grulke02}, where the role of large-scale fluctuation
structures has been highlighted. 
The role of parallel electron dynamics and Alfv\'en waves for
coherent drift modes and drift wave turbulence have been studied in a
collisionality dominated high-density helicon plasma \cite{grulke07}. 
Measurements of the phase coupling between spectral components of interchange
unstable drift waves at different frequencies in a basic toroidal magnetic
field configuration have indicated that the transition from a coherent to a
turbulent spectrum is mainly due to three-wave interaction processes
\cite{poli07}. 

The competition between drift wave and interchange physics in ExB 
drift turbulence has been studied computationaly in tokamak geometry with
respect to the linear and nonlinear mode structure by Scott \cite{scott05}. 
A quite remarkable aspect of fully developed drift wave turbulence in a
sheared magnetic field lying in closed surfaces is its strong nonlinear
character, which can be self-sustaining even in the absence of linear
instabilities \cite{scott-prl}. This situation of self-sustained plasma
turbulence does not have any analogy in neutral fluid dynamics and, as shown
in numerical simulations by Scott, is mostly applicable to tokamak edge
turbulence, where linear forcing is low enough so that the nonlinear physics
can efficiently operate \cite{scott02}. 

\section{Drift-Alfv\'en turbulence simulations for fusion plasmas}

The model DALF3 by Scott \cite{scott02}, in the cold ion approximation without
gyrofluid FLR corrections, represents the four field version of the 
dissipative drift-Alfv\'en equations, with disturbances  (denoted by the
tilde) in the ExB vorticity 
$\tilde\Omega$, electron pressure $\tilde p_e$, parallel current $\tilde
J_\parallel$, and parallel ion velocity $\tilde u_\parallel$ as
dependent variables. The equations are derived 
under gyro/drift ordering, in a three dimensional globally consistent
flux tube geometry  \cite{scott98,scott01}, and appear (in cgs units as used
in the references) as
\begin{eqnarray} \label{e:eqvor}
{n M_i c^2 \over B^2} \left( \partial_t + {\bf v}_E\cdot\bnabla
	\right) \tilde\Omega  
	&=& \nabla_\parallel \tilde j_\parallel  
	- {\cal K}(\tilde p_e), \\
{1 \over c} \partial_t \tilde A_{\parallel} 
	+ {m_e \over n_e e^2} \left( \partial_t + {\bf v}_E\cdot\bnabla
	\right) \tilde j_{\parallel}
	&=& {1 \over ne} \nabla_{\parallel} (p_e+\tilde p_e )
	- \nabla_{\parallel} \tilde\phi  
	- \eta_{\parallel} \tilde j_{\parallel},   \\
\left( \partial_t + {\bf v}_E\cdot\bnabla \right) (p_e+\tilde p_e) 
	&=& {T_e \over e} \nabla_{\parallel}\tilde j_{\parallel}  
	- p_e \nabla_{\parallel} \tilde u_\parallel
	+ p_e {\cal K} (\tilde \phi)
	- {T_e \over e} {\cal K} (\tilde p_e), \nonumber \\ \\
 n M_i \left( \partial_t + {\bf v}_E\cdot\bnabla \right)\tilde u_\parallel &=& -
 \nabla_\parallel (p_e+\tilde p_e ), 
\end{eqnarray}
with the parallel magnetic potential $\tilde A_\parallel$ given by
$\tilde j_\parallel= - (c / 4 \pi) \nabla_{\perp}^2 A_\parallel $
through Ampere's law, and the vorticity 
$\tilde\Omega=\nabla_\perp^2\tilde\phi.$
Here, $\eta_\parallel$ is the Braginskii parallel resistivity,
$m_e$ and $M_i$ are the electron and ion masses, $n$ is the electron
(and ion) density, and $T_e$ is the electron temperature with pressure
$p_e=n T_e$.  
The dynamical character of the system is further determined by a set of
parameters characterising the relative role of dissipative, inertial and
electromagnetic effects in addition to the driving by gradients of
density and temperature. 

The flux surface geometry of a tokamak enters into the fluid and gyrofluid
equations 
via the curvilinear generalisation of differentiation operators and via
inhomogeneity of the magnetic field strength $B$. The different scales of
equilibrium and fluctuations parallel and perpendicular to the magnetic
field motivate the use of field aligned flux coordinates.
The differential operators in the field aligned frame are the parallel
gradient  
\begin{equation}
\nabla_\parallel = (1/B) ({\bf B} + \tilde{\bf B}_\perp) \cdot \bnabla,
\end{equation}
with magnetic field disturbances $\tilde{\bf B}_\perp = (-1/B) {\bf B}
\times \bnabla \tilde A_\parallel$ as additional nonlinearities,
the perpendicular Laplacian
\begin{equation}
\nabla_\perp^2=\bnabla\cdot[(-1/B^2){\bf B}\times({\bf B}\times\bnabla)],
\end{equation}
and the curvature operator
\begin{equation}
{\cal K}=\bnabla\cdot[(c/B^2){\bf B}\times\bnabla)].
\end{equation}

The DALF equations constitute the most basic model containing the
principal interactions of dissipative drift wave physics in a general closed
magnetic flux surface geometry.  The drift wave coupling effect is
described by $\nabla_\parallel$ acting upon 
$\tilde p_e/p_e-e\tilde\phi/T_e$ and $\tilde J_\parallel$, 
while interchange forcing is described by ${\cal K}$ 
acting upon $\tilde p_e$ and $\tilde\phi$ \cite{scott97a}.  
In the case of tokamak edge turbulence, the drift wave effect is qualitatively
more important \cite{scott02}, while the most important role for ${\cal K}$ is
to regulate the zonal flows \cite{scott03pla}. 
Detailed accounts on the role of magnetic field geometry shape in tokamaks
and stellarators on plasma edge turbulence can be found in
Refs.~\cite{akpop06,akjpp06} and \cite{akppcf00,jenko02}, in particular with
respect to effects of magnetic field line shear \cite{akprl03} and curvature
\cite{akpop05}.  

An example for typical experimental parameters are those of the ASDEX Upgrade
(AUG) edge pedestal plasmas in L mode near to the L-H transition for Deuterium
ions with $M_i =  M_D$:  
electron density $n_e = 3 \cdot 10^{13} \mbox{cm}^{-3}$, temperatures $T_e =
T_i = 70$ eV, magnetic field strength $B = 2.5$ T, major radius $R = 165$ cm,
perpendicular gradient length $L_{\perp} = 4.25$ cm, and safety factor $q =
3.5$. 

The dynamical character of the DALF/GEM system is determined by a set of 
parameters characterising the relative role of dissipative, inertial and
electromagnetic effects in addition to the driving by gradients of
density and temperature.  In particular, for the above experimental
values, these are collisionality $C = 0.51 \hat\epsilon (\nu_e
L_\perp / c_s) (m_e/M_i)=5$, magnetic induction $\hat \beta = \hat\epsilon (4
\pi p_e / B^2)=1$, electron inertia $\hat \mu = \hat\epsilon (m_e/M_i) =5$
and ion inertia $\hat\epsilon = (qR / L_{\perp})^2=18350$.

The normalised values are similar in edge plasmas of other large
tokamaks like JET. The parameters can be partially obtained even by smaller
devices like the torsatron TJ-K at University of Stuttgart \cite{tjk,tjk-sim},
which therefore provides ideal test situations for comparison between
simulations and the experiment \cite{tjk07}. 

\begin{figure}
\includegraphics[width=11.5cm]{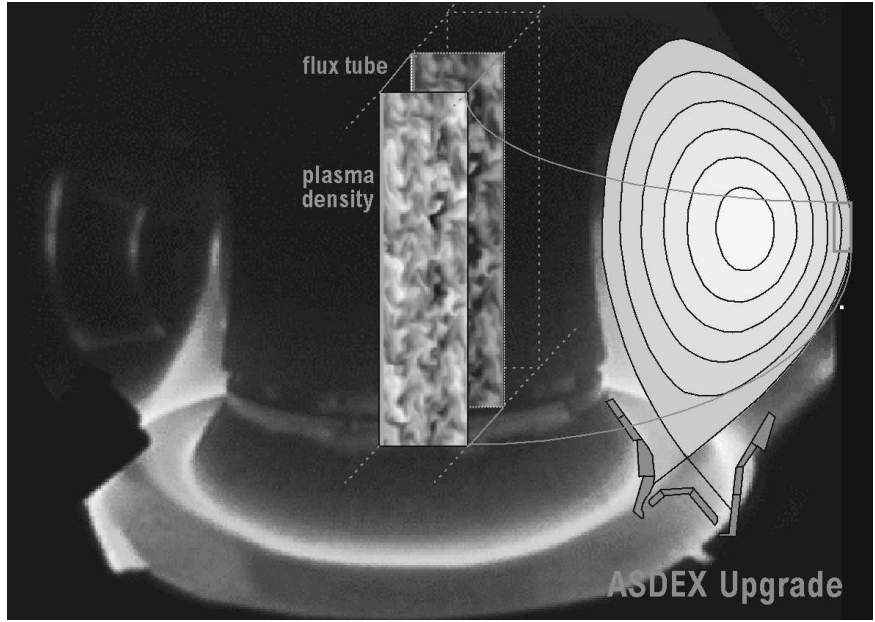}
\caption{\label{f:aug} \sl
Computation of plasma edge turbulence in the magnetic field of a divertor
tokamak fusion experiment, using the DALF3 model as described in the text
[Background figure: ASDEX Upgrade, Max-Planck-Institute for Plasma Physics].
}
\end{figure}

A review and introduction on drift wave theory in inhomogeneous
magnetised plasmas has been presented by Horton in Ref.~\cite{horton-rev},
although its main emphasis is placed on linear dynamics. 
An excellent introduction and overview on turbulence in magnetised plasma and
its nonlinear properties by Scott can be found in Ref.~\cite{slnp-scott}, and
a very detailed survey on drift wave theory with emphasis on the plasma
edge is given by Scott in Refs.~\cite{habil-scott,scott07}.

However, no tokamak edge turbulence simulation has yet reproduced the
important threshold transition to the high confinement mode known from
experimental fusion plasma operation. 
The possibility to obtain a confinement transition within first principle
computations of edge turbulence will have to be studied with models
that at least include full temperature dynamics, realistic flux surface
geometry, global profile evolution including the equilibrium, and a coupling
of edge and SOL regions with realistic sheath boundary conditions. In addition
the model still has to maintain sufficient grid resolution, grid deformation
mitigation, and energy plus enstrophy conservation in the vortex/flow system.

Such ,,integrated'' fusion plasma turbulence simulation codes are currently
under development. 
The necessary computing power to simulate the extended physics models and
computation domains is going to be available within the next years.
This may facilitate international activities (for example within the European
Task Force on Integrated Tokamak Modelling) towards a
,,computational'' tokamak plasma with a first-principles treatment of both
transport and equilibrium across the whole cross section. 
The objective of this extensive project in Computational Plasma Physics is to
provide the means for a direct comparison between our theoretical
understanding with the emerging burning-plasma physics of the next large
international fusion experiment ITER. 

\section*{Acknowledgements}
This work was supported by the Austrian Science Fund FWF under contract
P18760-N16, and by the European Communities under the Contract of
Associations between Euratom and the Austrian Academy of Sciences and
carried out within the framework of the European Fusion Development
Agreement. The views and opinions herein do not necessarily reflect those of
the European Commission.

\end{document}